# Residue mobility has decreased during protein evolution.

Charudatta Navare, Anirban Banerji*

Bioinformatics Centre, University of Pune, Pune 411007, Maharashtra, India

Contact address: **anirbanab@gmail.com**

## Abstract

Upon studying the B-Factors of all the atoms of all non-redundant proteins belonging to 76 most commonly found structural domains of all four major structural classes, it was found that the residue mobility has decreased during the course of evolution. Though increased residue-flexibility was preferred in the early stages of protein structure evolution, less flexibility is preferred in the medieval and recent stages. GLU is found to be the most flexible residue while VAL recorded to have the least flexibility. General trends in decrement of B-Factors conformed to the general trend in the order of emergence of protein structural domains. Decrement of B-Factor is observed to be most decisive (monotonic and uniform) for VAL, while evolution of CYS and LYS flexibility is found to be most skewed. Barring CYS, flexibility of all the residues is found to have increased during evolution of α/β folds, however flexibility of all the residues (barring CYS) is found to have decreased during evolution of all-β folds. Only in α/β folds the tendency of preferring higher residue mobility could be observed, neither α+β, nor all-α nor all-β folds were found to support higher residue-mobility. In all the structural classes, the effect of evolutionary constraint on polar residues is found to follow an exactly identical trend as that on hydrophobic residues, only the extent of these effects are found to be different. Though protein size is found to be decreasing during evolution, residue mobility of proteins belonging to ancient and old structural domains showed strong positive dependency upon protein size, however for medieval and recent domains such dependency vanished. It is found that to optimize residue fluctuations, α/β class of proteins are subjected to more stringent evolutionary constraints.

## 1. Introduction:

Proteins are not classical solids (Allen et al., 2004; Banerji and Ghosh, 2009a,b). The 'compact object description' of proteins (characterized by small-amplitude vibrations and by a low-frequency Debye density of states) cannot account for their non-idealistic behaviors (De Leeuw et al., 2009; Reuveni, 2008; Banerji and Ghosh, 2011). Indeed, the non-invariance of distance between any two atoms ($(\vec{r_a} - \vec{r_b}) \neq constant$) in any biologically functional protein can easily be verified with the simplest of computer programs. On one hand, proteins attempt to maintain the structure of their native fold thermally stable. On the other hand, such native fold template needs to accommodate large amplitude conformational changes that allow appropriate functioning of the protein (Karplus and McCammon, 1983; Bahar et al., 1998; Henzler-Wildman et al., 2007; de Leeuw et al., 2009). These two properties, viz. structural flexibility and structural rigidity, are not independent of each other; instead it has been found that fluctuations in densely packed regions manipulate the motion of flexible parts of proteins (Eisenmesser et al., 2005; Huang and Montelione, 2005). The Debye-Waller factor (alternatively called the 'B-factor', or the 'temperature-factor') is a reliable construct to measure residue flexibility in a local scale.

Globular proteins embody a wide spectrum of internal motions; furthermore, the time scale of protein mobility is very wide too (the fastest vibrations and motions requiring only $10^{-14}$ to $10^{-13}$ seconds). B-factors provide an indirect way to quantify the local mobility by measuring atomic displacement of protein residue atoms. B-factors quantify the decrement of intensity in diffraction caused by the dynamic disorder (owing to temperature-dependent vibration of the atoms) and the static disorder (related to the orientation of the molecule) (Schlessinger and Rost, 2005). Magnitude of B-factor of an atom is obtained by calculating: $8\pi^2 \langle u^2 \rangle$, where $\langle u^2 \rangle$ denotes the average of the mean square atomic displacements ($\text{Å}^2$) along the three coordinate axes, and is given by: $[(u_x^2 + u_y^2 + u_z^2)/3]$.

The average mean square atomic displacements quantify the fluctuation of an atom about its average position and thereby provide a way to quantify the local mobility in proteins. Since local mobility is prerequisite to ensure local structural flexibility, a systematic analysis of protein B-factors provides a reliable way to investigate the structural flexibility in a protein (Vihinen et al., 1994, and references therein). High magnitude of B-factor indicates high mobility (and therefore, high flexibility) of individual atoms and residues, whereas low magnitudes of B-factors imply the presence of structural rigidity. The B-factors provide an alternative way to obtain information about the relative vibrational motion of different parts of a protein structure. Thus B-factors provide a rich source of information about the dynamics and flexibility in a protein.

One can estimate the sheer importance of structural flexibility of protein residues by observing that protein internal motion or flexibility is highly correlated with protein functions such as catalysis and allostery (Teilum et al., 2011). More generally, the structural flexibility has been associated with various other biological processes, notably in processes like molecular recognition and catalytic activity (Carr et al., 1997; Teague, 2003; Yuan et al., 2003; Daniel et al., 2003; Yuan et al., 2005). While studies on role of structural flexibility and mobility on protein functions are numerous, we ask a different question in this work; *has evolution preferred structural flexibility?* Intuitively, given the available pool of literature about benefits of more structural flexibility and mobility of protein residues, the answer to the aforementioned question seems to be an affirmative "yes"; - but is it really so? More importantly, how to quantify the universal patterns in the extents to which evolution has preferred or disallowed structural flexibility? – Even though works on residual mobility, flexibility, and dynamics are (practically) innumerable, finding an unambiguous answer to this simple question is not so easy. In the present work, we attempt to find a general answer to the aforementioned question by studying the cumulative B-factor profiles in statistically significant number of non-redundant proteins populating 75 common structural folds, which are sorted with evolutionary scale and which are distributed across four major structural classes of globular proteins.

A comprehensive description of the structural relationships between known protein structures can be obtained from the SCOP (Structural Class of Proteins) database (Murzin et al., 1995). SCOP presents information about the hierarchical organization of proteins in respective domain structures. The classification on the first level of the hierarchy is commonly known as the protein structural class, while the second level classifies proteins into folds. The evolutionary scale index magnitude could be assigned to many of the SCOP folds (Caetano-Anollés and Caetano-Anollés, 2003). Hence, an exhaustive work on characteristics of evolutionarily-sorted set of SCOP folds present an ideal way to investigate evolution of a protein property across various structural domains distributed under major SCOP classes. In the present work, we attempted to probe the trends in evolution of local flexibility (as quantified by B-Factors) across structural domains along the evolutionary timeline.

A relevant point should be clarified here. The experimentally determined magnitude of B-factor is not a permanent parameter; it depends on the choice of refinement procedures (Tronrud, 1996), the resolution of protein structure, crystal contacts (Sheriff et al., 1985). However, these concerns assume importance when one attempts to compare the B-factors from different structures. Since the present work concentrates on the mean behavior of B-factors of statistically significant number of proteins in various SCOP folds (without attempting to compare B-factor of one protein with that of another), the aforementioned influencing factors can safely be regarded as inconsequential for the present study.

## 2. Materials and Methods:

**2.1:** *Materials:* Protein crystal structures with higher resolution are endowed with more reliable B-factor magnitudes. Therefore, the present study was restricted only to structures with resolutions below 2.5 Å. Since within this resolution limit, the correlation between resolution and the log of the mean diffraction intensity is (almost) linear (Blow, 2002), the B-factor assignment in these structures could be assumed to be trustworthy. It was further ensured that the R-factor of each of these structures is less than 0.2. Additionally, it was made sure that none of the proteins considered contained any "disordered" regions (as defined in (Dunker et al., 2002)) in them. All the non-redundant proteins in Protein Data Bank (Berman et al., 2003) belonging to 80 common structural folds (20 from each structural class) were chosen for investigation initially. It was necessary for analysis that all the folds considered here are assigned their evolutionary scale index (Caetano-Anollés and Caetano-Anollés, 2003), which describes their emergence in the course of protein structure evolution. The assortment of structural folds thus obtained contained all the 'most populous' SCOP folds enlisted (as training-set and test-set) in a recent work (Chen and Kurgan, 2007). However some of the proteins from folds Nucleoplasmin like VP viral coat and capsid proteins (all-β class), P loop containing nucleoside triphosphate hydrolases (α/β class), Thioredoxin fold (α/β class), and Knottins small inhibitors toxins lectins ('Small proteins' class) – failed to match certain statistical criteria; whereby these folds were not considered for the main analyses.

**2.2:** *Methods:* Methodology for this work is absolutely simple and it follows a linear scheme. The atomic displacement parameter for each atom of each residue was averaged to find the residual B-factor. Using these values, average B-factor of every amino acid in a protein was calculated. The proteins are collated according to their SCOP folds, whereby SCOP fold-specific B-factor profile could be generated by averaging the B-factors of the resident proteins. The SCOP fold-specific B-factors were averaged to find the SCOP-class specific B-factor. Since, evolutionary scale magnitude could be associated with all the SCOP-folds considered for the present work, distribution of B-factors in structural domains could be studied from an evolutionary perspective.

**3. Results and Discussion:**
**3.1: Reduction of residual mobility, from purely the perspective of structural organization**
Upon phylogenetic analyses a previous paper (Caetano-Anollés and Caetano-Anollés, 2003) has established that structural classes of globular proteins appeared in the course of evolution in the defined order: α/β→–α+β→–all-α→–all-β. To view the general nature of residue mobility across four major SCOP classes, the residue B-factors are plotted in this very order.

The general trend of residue B-Factors(BF) show that mobility of every residues has slowed down in varying extent from α/β to all-β structural class, along the α/β→–α+β→–all-α→–all-β pathway of emergence of structural classes. However, though the decrement of residual mobility is unmistakable, quantum of decrement in BF between any two states of the aforementioned pathway, varied; whereby certain residues display a monotonic nature slowing down of mobility, certain others don't. Based on the monotonicity of trends (**Fig-1**) shows that THR, VAL, PHE, PRO, SER have slowed down most uniformly at each stage of transition along the α/β→–α+β→–all-α→–all-β pathway of structural class emergence, whereas ARG, ASP, GLY, HIS, MET have lost mobility but not uniformly. The non-uniformity in BF decrement is even more pronounced for ASN, GLU, ALA, LEU, ILE, TYR, TRP - where all-α and all-β BFs shared almost the same magnitude. In contrast to all these 17 trends, the BF decrement trends for CYS, LYS, GLN showed non-monotonicity, where all-α BFs registered lower magnitude than all-β BFs. The resultant magnitude of BFs of all the residues is presented below in **Table-1**.

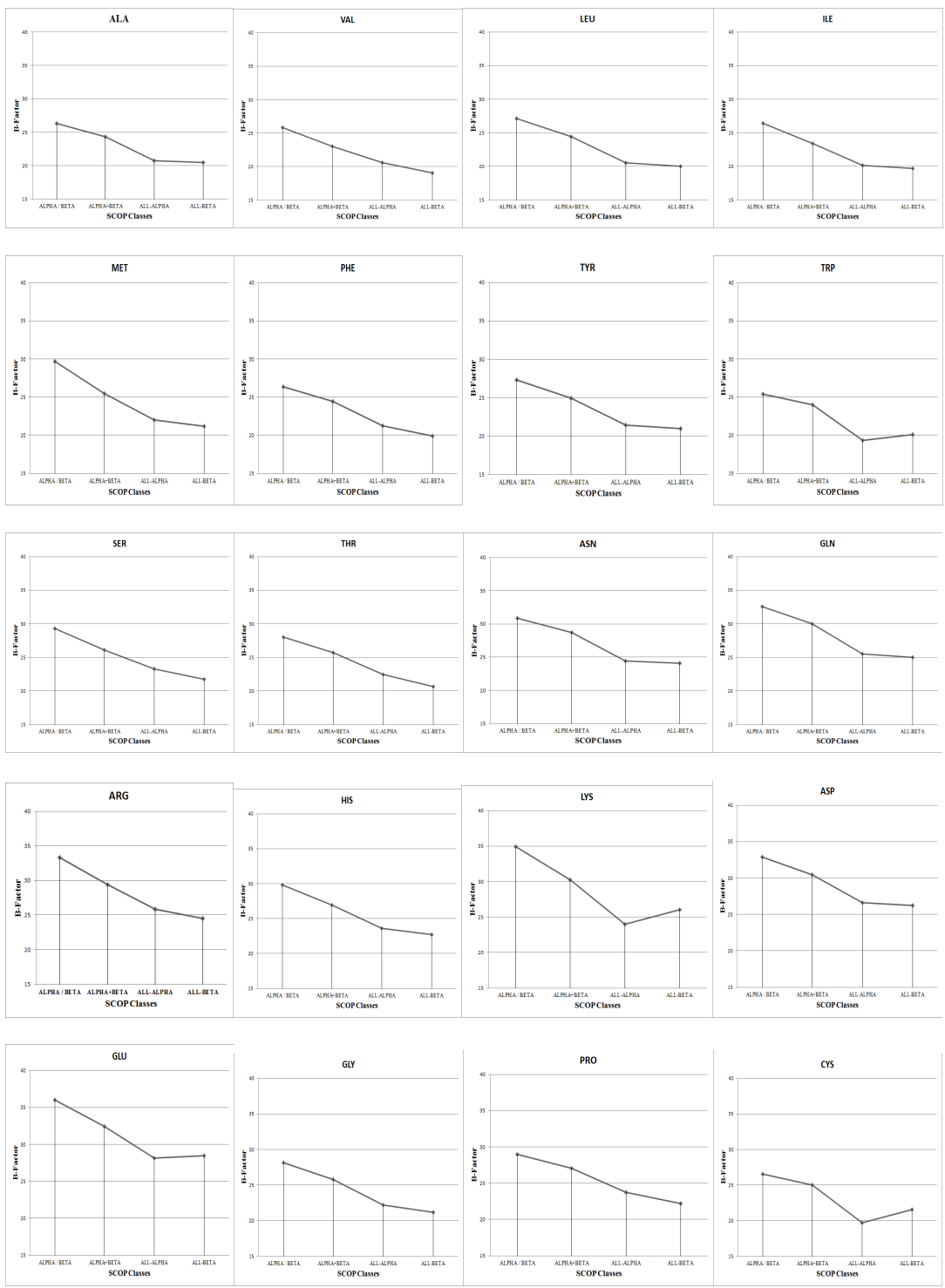

**Figure-1: structural class-specific B-factor for each residue**

**Table-1: Mobility of residues**

| RES | B-Factor Mean | B-Factor Standard-Deviation |
|---|---|---|
| ALA | 22.94426667 | 5.566002538 |
| ARG | 28.20613333 | 7.227302568 |
| ASN | 26.91666667 | 6.549689147 |
| ASP | 28.93453333 | 6.815601794 |
| CYS | 23.3008 | 7.153394159 |
| GLN | 28.2044 | 7.143064911 |
| GLU | 31.25333333 | 7.293738769 |
| GLY | 24.24933333 | 6.32714088 |
| HIS | 25.71813333 | 6.522764025 |
| ILE | 22.34506667 | 5.702676084 |
| LEU | 22.98333333 | 6.261251566 |
| LYS | 28.84093333 | 8.020764467 |
| MET | 24.5444 | 7.291671154 |
| PHE | 22.83533333 | 5.816291271 |
| PRO | 25.40653333 | 6.419090508 |
| SER | 25.00346667 | 6.907573887 |
| THR | 24.09693333 | 6.215560242 |
| TRP | 22.164 | 6.616496378 |
| TYR | 23.59333333 | 5.975177558 |
| VAL | 22.02773333 | 5.737354738 |

(A structural class-specific break-up of **Table-1** is presented in **Fig.-1**.)

But how consistent are these trends? Residue mobility in α/β proteins is found to be maximum, does that imply that variability in residue mobility is maximum in α/β proteins? Similarly, do the **Fig.-1** trends imply that residues in all-β proteins are uniformly restrained, since the mean profile of their mobility is found to be least in all-β proteins? – These questions can be addressed by analyzing the standard-deviations for each of the residues across SCOP classes, which is presented in **Table-2**.

**Table-2: Standard Deviations in residue mobility in major structural classes**

| RES | α/β | α+β | All-α | All-β |
|-----|-----|-----|-------|-------|
| ALA | 2.823789694 | 7.282639501 | 5.83226959 | 5.822978506 |
| ARG | 3.22108723 | 9.12202114 | 7.124224937 | 6.872684758 |
| ASN | 3.476003687 | 8.229356454 | 7.539019451 | 6.082314721 |
| ASP | 3.988486207 | 8.825897184 | 7.084116707 | 6.678792122 |
| CYS | 3.375173876 | 8.492459597 | 7.972657919 | 7.177667412 |
| GLN | 4.516278798 | 8.94262259 | 7.415272721 | 6.845117661 |
| GLU | 3.380717331 | 9.606258109 | 7.698263069 | 7.511216525 |
| GLY | 3.327529506 | 7.884704652 | 6.942261673 | 6.292527293 |
| HIS | 3.075780415 | 8.145479595 | 7.434153617 | 6.4346996 |
| ILE | 2.921401367 | 7.252729197 | 6.103508125 | 5.726933045 |
| LEU | 2.897687212 | 7.627730833 | 6.301448662 | 6.156541472 |
| LYS | 3.029449841 | 9.941766669 | 8.40506157 | 7.873231447 |
| MET | 5.099828428 | 8.821050899 | 8.574871483 | 6.606423728 |
| PHE | 3.01235148 | 7.271933093 | 6.169883845 | 5.730422484 |
| PRO | 2.938737669 | 8.067321931 | 7.534372347 | 6.217165436 |
| SER | 3.711223791 | 8.434749821 | 7.638679642 | 6.57951214 |
| THR | 3.284985898 | 7.841353979 | 6.057829015 | 6.232597245 |
| TRP | 3.555887367 | 7.658686995 | 7.055574156 | 7.336125089 |
| TYR | 3.422276504 | 7.526142735 | 6.56920072 | 5.574570647 |
| VAL | 2.637800443 | 7.041490254 | 6.120662507 | 5.689308417 |

The striking observation that can be made from **Table-2** is that though the mobility of residues in α/β class of proteins is found to be maximum (**Fig.-1**), the variability around their high mobility profile is the minimum therein (mean of the standard deviations of residues in α/β is found to be merely 3.00, significantly lower than that in α+β, all-α and all-β proteins, assuming magnitudes 6.42, 6.86 and 6.19 respectively). This demonstrates that in α/β proteins residue mobility is not only maximum, such trend of maximum magnitudes are most consistent when compared against the same observed in other structural classes. Implications of these findings are discussed in details in the section 'The unique set of evolutionary and structural constraints on α/β class of proteins'.

### 3.2. Reduction of residual mobility, under purely evolutionary constraints

The suit of figures in **Fig-1**, though informative, fails to throw much light on the evolutionary dynamics of reduction of residue mobility in four different contexts of structural organization. We start with the broad quantification of the extent of reduction of mobility for each residue during evolution, without considering the structural classes that these residues belonged to. Such data (presented in **Table-3**) demonstrates that most of the residues were made to reduce their mobility in similar manner, mean of correlation coefficients between evolutionary scale index for the folds and fold-specific (mean) BF for residues could be noted to be -0.31. In overall terms, mobility of LEU has decreased maximally during the evolutionary journey, while CYS suffered minimal losses to its mobility. Case of CYS's mobility is of immense interest and it will be discussed in details later.

**Table-3: Overall profile of B-Factor evolution for all the residues**

| RES | Correlation Coefficient between B-Factor and Evolutionary Scale |
|---|---|
| ALA | -0.358466883 |
| ARG | -0.324228288 |
| ASN | -0.235078745 |
| ASP | -0.313033701 |
| CYS | -0.162080629 |
| GLN | -0.315775849 |
| GLU | -0.327206084 |
| GLY | -0.31913302 |
| HIS | -0.320857506 |
| ILE | -0.327058177 |
| LEU | -0.376054666 |
| LYS | -0.263067463 |
| MET | -0.256678611 |
| PHE | -0.333394941 |
| PRO | -0.291940801 |
| SER | -0.280601637 |
| THR | -0.341403244 |
| TRP | -0.344261285 |
| TYR | -0.310619895 |
| VAL | -0.361705277 |

Though informative, **Table-3** depicts only the cumulative degree to which residue nobilities were made to decrease with evolution. But evolution of protein structure cannot be assumed to have a linear and continuous nature with equally-spaced intervals. Thus to answer the question 'does evolution prefer mobility?' one needs to analyze the increment or decrement of mobility in adequately resolved evolutionary phases, without paying attention to which structural classes the folds belong to. Such a study of evolutionary dynamics of mobility can be found in **Fig-2(A-D)**, which presents the trends in evolution of mobility in 'ancient folds' (folds with evolutionary scale < 0.1), 'old folds' (folds for which 0.1 < evolutionary scale < 0.2), 'medieval folds' (folds for which 0.2 < evolutionary scale < 0.5), and 'recent folds' (folds with evolutionary scale > 0.5), respectively. As demonstrated unambiguously, proteins belonging to 'ancient folds' were encouraged to promote more and more residue mobility, as a result the total mobility of the folds increased steadily with passage of evolution (**Fig.-2A**). But such increase of net mobility was perhaps not entirely beneficial; hence in the next lot of emerging folds, that is, in the 'old folds', mobility was made to decrease sharply (described by the marked slope of trend-line in **Fig.-2B**). But then again, such extent of strict restraint on mobility was found to be not beneficial either; and hence, during the next lot emerging folds the mobility was encouraged to increase again, whereby the pronounced slope of **Fig.-2B** (that marked reduction of mobility in 'old folds' with evolutionary passage), were fixed to a complete horizontal trend-line in the case of 'medieval folds' (**Fig.-2C**). Thus, from the perspective of transition from old to medieval folds, the horizontal trend-line of (**Fig.-2C**) does not imply merely an invariant profile of mobility along evolutionary journey but signifies an increase of the same. Finally, during the emergence of the last lot of folds (viz., the 'recent folds'), evolution has chosen to decrease the mobility again (**Fig.-2D**), but not with the zeal with which it chose to restrain the mobility of old folds.

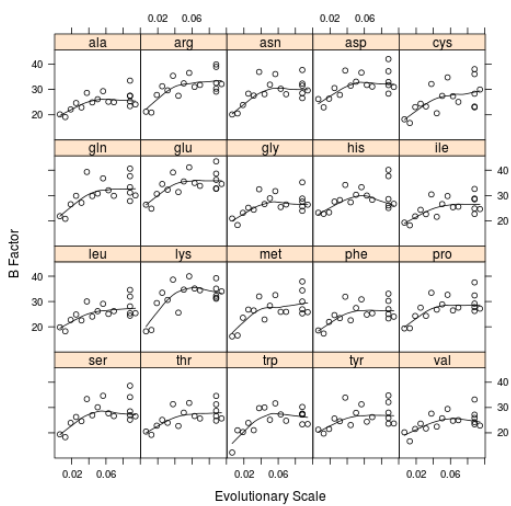

**Fig.-2.A**

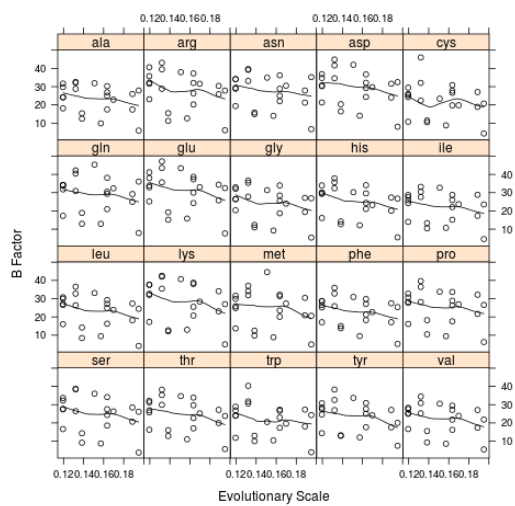

**Fig.-2.B**

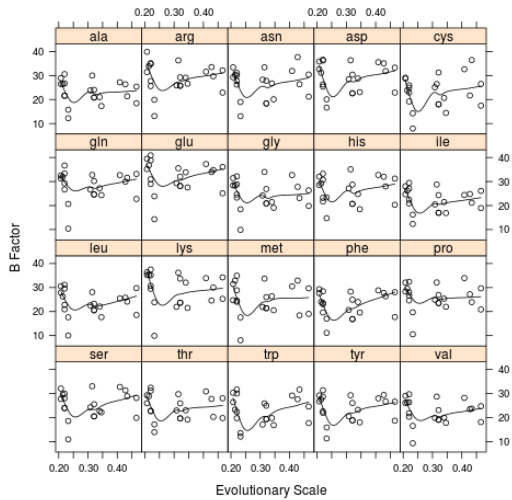

**Fig.-2.C**

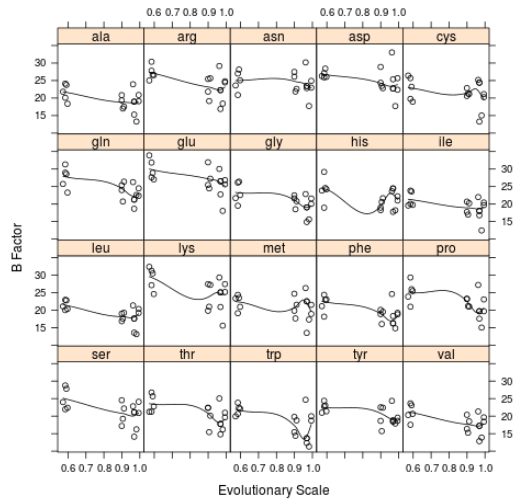

**Fig.-2.D**

**Figure 2A: trends in evolution of mobility in 'ancient folds'; Figure 2B: trends in evolution of mobility in 'old folds'; Figure 2C: trends in evolution of mobility in 'medieval folds'; Figure 2D: trends in evolution of mobility in 'recent folds'**

*Though emergence of structural domains followed the sequence: α/β→–α+β→–all-α→–all-β, one should not confuse between order of emergence of folds (a structural entity) with evolutionary scale index (derived from evolutionary considerations on sequences). Therefore, one should not wrongly assume that all the 'ancient folds' (folds with evolutionary scale < 0.1) are α/β or α+β domains. The classification 'ancient', 'old', 'medieval' and 'recent' are made only to investigate the evolutionary dynamics in better resolution in a reduced noise. Thus, though all-α and all-β domains emerged last among the structural classes, the 'ancient' folds included all-α domains like 'DNA-RNA binding 3 helical bundle'(evolutionary scale index: 0.006289308), 'Alpha Alpha superhelix'(evolutionary scale index: 0.088050315); and all-β domains like Oligonucleotide/Oligosaccharide-binding fold (OB-fold, that is found in all three kingdoms)(evolutionary scale index: 0.044025157), and 'Double Stranded Beta Helix'(evolutionary scale index: 0.088050315).*

## 3.3: Evolution of residual mobility, under both evolutionary and structural constraints

Findings enlisted in **Table-3**, however fails to throw much light on the nature of BF evolution under the particular stability constraints enforced by each of structural classes. This result is presented in Table-4, which consists of correlation coefficients between BF of each of the residues in each of the folds belonging to any of the four SCOP classes, and evolutionary scale index of these folds. Correlation coefficients thus obtained tells us how, during evolutionary journey, the mobility of each residue was made to increase or decrease, when the residue is subjected to reside in any of the four structural states. The most remarkable information one derives from Table-4 is that the collective effect of evolutionary and structural constraints ensured that the mobility of residues belonging to α/β class proteins is increased during evolution. As an effect of this convergence, barring CYS, for all the residues (hydrophobic, hydrophilic alike) positive correlation coefficients could be obtained.

Mobility profiles in α+β, all-α and all-β classes were found to be not so linear. For as many as 13 out of 20 residues in all-α class of proteins, residual mobility is found to be (absolutely) independent of evolutionary status. As mentioned earlier, evolution of mobility of CYS was found to be conspicuously different than that of any others, the case of CYS will be discussed later.

**Table-4: Evolution of residual mobility under both evolutionary and structural class-specific constraints.**

| RES | α/β | α+β | All-α | All-β |
|---|---|---|---|---|
| ALA | 0.498775642 | -0.310596554 | -0.241006 | -0.208275 |
| ARG | 0.591989803 | -0.228540192 | -0.11724 | -0.136707 |
| ASN | 0.302759021 | -0.145306482 | -0.043059 | -0.043059 |
| ASP | 0.435860803 | -0.230815584 | -0.236682 | -0.144871 |
| CYS | -0.099075831 | -0.171551628 | -0.018671 | 0.0203803 |
| GLN | 0.235471796 | -0.247611489 | 0.0017586 | -0.215083 |
| GLU | 0.334314109 | -0.286140985 | -0.211877 | -0.125911 |
| GLY | 0.544532915 | -0.250265659 | -0.081453 | -0.193505 |
| HIS | 0.516390788 | -0.362413645 | -0.018365 | -0.201466 |
| ILE | 0.380986538 | -0.249461833 | -0.199204 | -0.110311 |
| LEU | 0.470812927 | -0.317462531 | -0.2138 | -0.206182 |
| LYS | 0.278840837 | -0.200517267 | -0.045243 | -0.048114 |
| MET | 0.523991105 | -0.184265023 | 0.0045418 | -0.15317 |
| PHE | 0.148969862 | -0.244098466 | -0.075031 | -0.175741 |
| PRO | 0.429740248 | -0.200548694 | -0.092139 | -0.137482 |
| SER | 0.391550985 | -0.28357835 | -0.050354 | -0.060787 |
| THR | 0.452245539 | -0.21593194 | -0.084569 | -0.224104 |
| TRP | 0.234210592 | -0.330327195 | 0.0288267 | -0.321829 |
| TYR | 0.314303973 | -0.253818245 | -0.02053 | -0.197157 |
| VAL | 0.53889197 | -0.288807298 | -0.18701 | -0.190576 |

### 3.3: Evolution's way of managing mobility of hydrophobic and hydrophilic residues

It is commonly believed that atoms, when part of hydrophobic residues possess lower BFs as compared to the case when they are part of hydrophilic residues. Rationale for such an idea stems from the fact that the hydrophobic residues are typically found in the interior of the proteins; thus possibility of their atoms to experience fluctuation due to solvent-protein interactions will be low, whereby the BF of atoms belonging to hydrophobic residues will not be as high as the case when they would have constituted hydrophilic residues. – Though a logical expectation, this opinion does not answer the simple questions like:

    Do evolutionary constraints influence only the extent of mobility of hydrophobic and hydrophilic residues or do they compel the mobility of hydrophobic and hydrophilic residues to follow altogether different trends?

    Is there any uniformity in the way the structural (geometrical and/or mechanical) constraints of SCOP classes regulate residue mobility?

    Is there any difference the way in which any structural class manages the mobility of hydrophobic residues from that of hydrophilic residues?

    etc..

To find a unified and general answer to these questions, upon retrieving the SCOP-class-specific mean BF of strongly hydrophobic (VAL, ILE, LEU, MET, PHE, TRP and CYS) and strongly hydrophilic (ARG, LYS, ASP, GLU, ASN, GLN and HIS) residues, they were sorted with respect to evolutionary scale. Patterns obtained (**Fig-3**) demonstrate that the hydrophobic residues (aliphatic and aromatic alike) have not only low mobility but also low range of change in their BF magnitude, whereas for the hydrophilic residues, expectedly, this trend just reversed. However, the difference in evolution's way of handling the mobility of hydrophobic and hydrophilic residues could only be noted in the extent of residual mobility. As a result, for each of the four structural classes, the trends describing evolution of mobility of hydrophobic and hydrophilic residues registered identical patterns. For α/β class of proteins, the difference between mobility of hydrophilic and hydrophobic residues is found to be the maximum, 5.42Å$^2$; for α+β, all-α and all-β class of proteins, the difference in mobility settled down to lower range, 4.86Å$^2$, 4.27Å$^2$ (the least) and 4.66Å$^2$, respectively. But it will be wrong to interpret from these values that there is uniformity in the manner in which evolutionary constraints influence different structural classes to manage their residual mobility. A closer inspection of **Table-5** reveals that both hydrophilic and hydrophobic residues possess their maximum mobility in α/β class of proteins, then for both of them, the mobility reduces along α+β→–all-α→–all-β journey; – a finding that bears distinct resemblance to **Fig-1** results.

**Table-5: Difference hydrophobic and hydrophilic mobility in structural classes.**

|   | (Mean of) hydrophilic residue BF (Å$^2$) | (Mean of) hydrophobic residue BF (Å$^2$) | (Mean of) Differences |
|---|---|---|---|
| **α/β proteins** | 32.448 | 27.028 | 5.42 |
| **α+β proteins** | 29.292 | 24.427 | 4.86 |
| **All-α proteins** | 25.70 | 21.43 | 4.27 |
| **All-β proteins** | 24.86 | 20.20 | 4.66 |

Reduction of residue mobility, however, is far from being a uniform process. **Fig.-3** reveals that along evolutionary journey through α/β class of structural constraints, residue mobility did increase for both hydrophilic and hydrophobic residues. Interestingly, such evolutionary increase in residue mobility could also be detected for all-α class of proteins, for both hydrophilic and hydrophobic residues. Structural constraints of α+β and all-β proteins, however, did not permit residue mobility to increase during evolutionary journey.

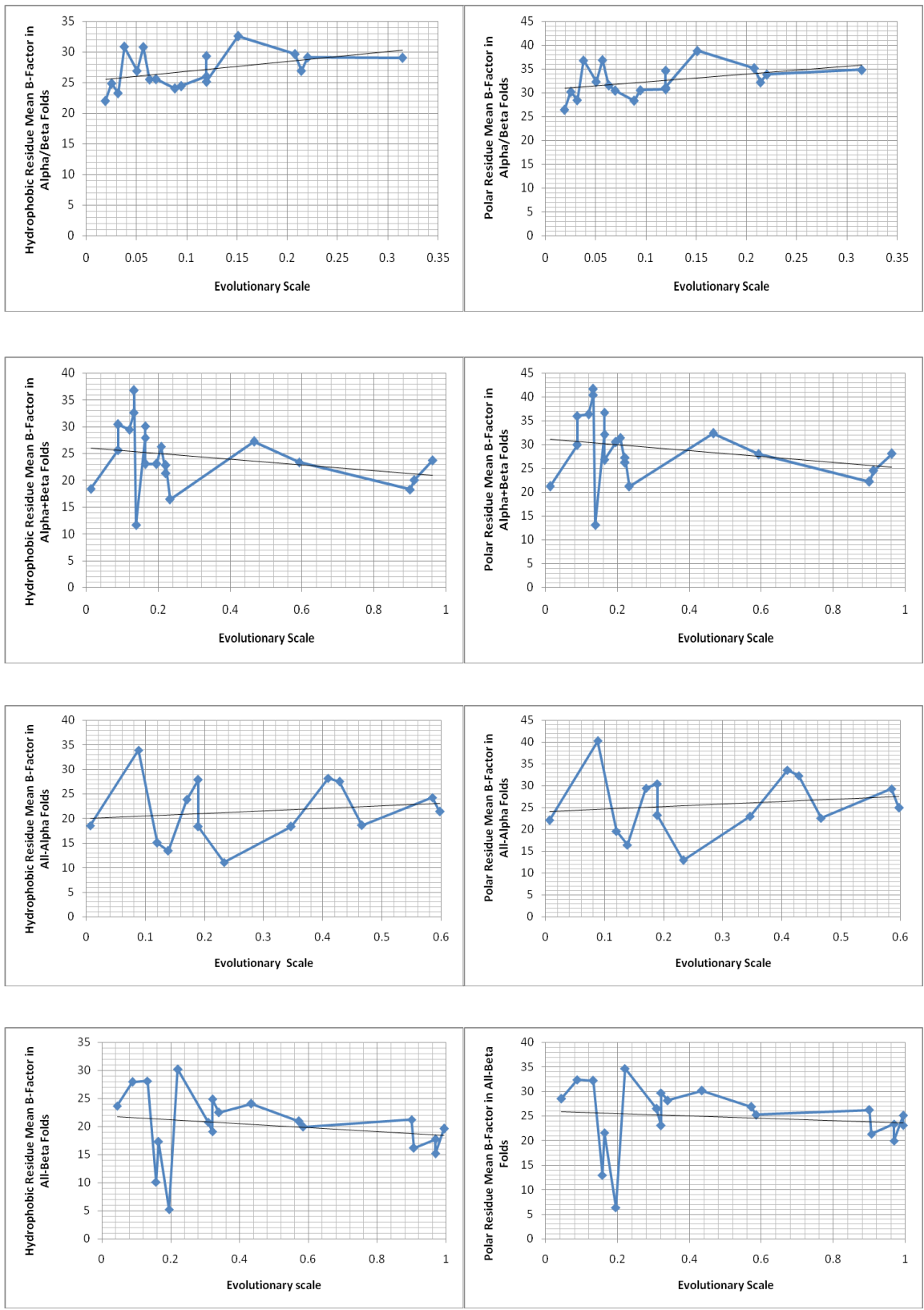

**Figure-3: Evolution of B-Factors of hydrophobic and hydrophilic residues in**

α/β, α+β, all-α and all-β structural classes.

### 3.3. Evolution of dependence of residue mobility on protein size in major SCOP classes:

Residue mobility depends crucially on protein size; therefore, studies on evolution of residue mobility needs to ascertain the nature of evolution of protein size. There are two facets of of analyzing the complex relationships in the way evolution has managed protein size (measured here with the number of residues and not by exact volumes) under the different organizational constraints (geometrical and mechanical) of protein structural classes. One, which studies trends in size-vs-mobility purely under evolutionary constraints, the other, that studies the same under both evolutionary as well as structural constraints. **Table-6** and **Table-7** present the trends respectively. Not unexpectedly, in the light of evolution, the commonly perceived notion of 'more size implies more fluctuation' was found to be inadequate to describe the complexity.

**Table-6: Probing protein size-versus-protein mobility in different phases of evolution.**

|  | Correlation Coefficient between Size(#residues) and Evolutionary Scale | Correlation Coefficient between Size(#residues) and residue fluctuations(BF) |
|---|---|---|
| **Entire span of evolutionary scale** | **-0.234** | **0.44** |
| **Proteins in 'Ancient folds'** | **0.386** | **0.233** |
| **Proteins in 'Old folds'** | **-0.268** | **0.697** |
| **Proteins in 'Medieval folds'** | **0.2118** | **0.059** |
| **Proteins in 'Recent folds'** | **0.4729** | **0.039** |

(As mentioned earlier, 'ancient folds' are defined as folds with evolutionary scale < 0.1, 'old folds' are folds for which 0.1 < evolutionary scale < 0.2, 'medieval folds' are folds for which 0.2 < evolutionary scale < 0.5, and 'recent folds' are folds with evolutionary scale > 0.5)

**Table-7: Probing protein size-versus-protein mobility in different structural classes**

| SCOP Class | Correlation Coefficient between Size(#residues) and Evolutionary Scale | Correlation Coefficient between Size(#residues) and residue fluctuations(BF) |
|---|---|---|
| **α/β proteins** | **-0.294** | **-0.396** |
| **α+β proteins** | **-0.263** | **0.590** |
| **All-α proteins** | **0.102** | **0.375** |
| **All-β proteins** | **0.155** | **0.248** |

Principles of protein physics tells that structural organizations (be it at the resolution of classes or of folds) essentially describe different geometric ways to arrange secondary (and super-secondary) structures and motifs so that, first, the mechanical stability of the constructed structure is ensured to be high, and second, adequate provisions are made to accommodate the necessary flexibility (which includes residue fluctuations too) of the structure – so that it can function (Frauenfelder et al., 1991; Rasmussen et al., 1992). Hence, size, viz. total number of residues accommodated in a protein and allowable extent of residue fluctuation, are both subjected to not only the evolutionary constraints (that is, which structure can function effectively consuming least energy) but also structural/stability constraints. Findings enlisted in **Table-7** suggest that during their evolutionary journey, proteins belonging to α/β and α+β structural classes are forced to reduce their size, but all-α and all-β class of proteins are encouraged to increase their size by small margin. When viewed from the perspective of order of emergence of structural classes, viz. α/β→–α+β→–all-α→–all-β, this data tends to suggest that though evolution was stringent to trim the proteins that belong to old structural classes, benefits (structural or functional or both) of more number of residues in a protein were recognized; whereby with the passage of time protein sizes were encouraged to increase.

– Such linear attempt to understand protein structure evolution, however, cannot explain the profile of dependence of net fluctuation profile of the residues on the total number of residues. For α/β proteins, not only the number of residues was made to reduce over time, but also the net fluctuation profile of the residues was made to reduce over time; but for α+β proteins, the effect of reduced number of residues was compensated by huge increase in residue mobility. As a result, during α/β→–α+β stage of structure evolution, correlation between net mobility of the residues and total number of residues grew from ~(-0.4) to ~(+0.6). But such overwhelming amplification of residue mobility was found to be non-beneficial and hence during α+β→–all-α transition correlation between net mobility of the residues and total number of residues shrunk from ~(+0.6) to ~(+0.38); before shrinking further (~(+0.38) to ~(+0.25)) during all-α→–all-β transition. Allowable extent of residue fluctuation in structural classes is dependent on spectrum of parameters involving protein physics and evolutionary constraints. Thus, only an exhaustive analysis of evolution of function-preserving mechanical stability of protein structural classes can explain the reported trends.

**3.4: The unique set of evolutionary and structural constraints on α/β class of proteins.**

Evolutionary constraints coupled with structural constraints have forced the α/β class of proteins to continuously reduce the number of residues that can be accommodated in a protein. Alongside this, the larger an α/β protein became, the less became its total residue mobility. Therefore, the expectation: 'more number of residues implies more total fluctuation' – fails completely in case evolution of α/β proteins. But to function effectively, a protein structure requires possessing flexibility and protein flexibility includes fluctuation profile of the constituent residues. Hence, one may hypothesize that to effectively function under the strict set of evolutionary and structural constraints, the α/β class of proteins had to adopt a strategy that maximizes the flexibility of the protein with their ever-shrinking population of residues. One may further hypothesize that to meet this very requirement the α/β class of proteins had to ensure that the fluctuation profile of every residue therein is increased to the maximum extent, which explains the consistent observation of high magnitudes of residue mobility for every residue that is part of α/β class of proteins (**Fig-1**). More importantly, as **Table-4** demonstrates, the residues that were made part of α/β proteins are encouraged to fluctuate more during evolutionary progress. Such a hypothesis nicely explains why the fluctuation profiles of both hydrophilic and hydrophobic residues are made to increase over the course of evolution (**Fig-3**); that is, as the (permitted) total number of residues is made to reduce over evolution, the ultimately accommodated residues (hydrophilic, hydrophobic alike) are made to fluctuate more to maximize the effectiveness of allowed flexibility.

**3.2: Effect of biophysical properties of residue on its B-Factor evolution**

One can estimate the local mobility of every atom in every residue from atomic displacement of this atom. B-factor of any atom of any protein residue provides an indirect means to quantify the local mobility of the atom. Though the structures of 20 amino acids are different from each other, many of these structures can be derived from some or the other structure(s) by substituting an atom or a group of atoms in an equivalent position. Analyses of BF of the related residues provides an alternative way to study the (comparative) effect of local mobility of one atom or a group of atoms on the mobility profile of the entire residue. How such contrasting nature of mobility between residues influences their physico-chemical properties in functioning proteins is an interesting question which the present section attempts to answer.

To ensure perfect generality, BF magnitude of every atom in every residue of every non-redundant protein in most-populous SCOP fold under all four major SCOP classes was considered. The final data thus obtained, though extensive, suffered from multiple levels of smoothening due to successive averaging. Furthermore, as can be observed from **Fig-1**, the windows of recorded residue-BFs at SCOP class level are found to vary within a small window between (22 Å$^2$ - 31Å$^2$)[**Table-1**]. Strikingly though, even after such numerous coarse-graining operations and even after being restricted to a narrow window of possible BF values, subtle yet definite differences in SCOP-class specific BF magnitudes could be observed. It is difficult to establish to what extent the residue mobility has been influenced by biophysical properties of the residue and to what extent the reverse happened. Nevertheless, few cases are discussed below.

### 3.2.1: *Comparing and contrasting B-Factor-based residue classifications with biochemistry-driven residue-classification.*

CYS (polar, uncharged) differs from SER (polar, uncharged) side-chain by a single atom; the sulfur of the thiol replaces the oxygen of the alcohol. At various other levels of biological organization the similarities between CYS and SER can be observed. (For example, the general mechanism of catalysis reaction by thiol proteinases can be observed to be the same as that of serine proteinases. CYS at position 25 in thiol proteinases papain can be observed to perform the same function as SER at position 195 in trypsin, chymotrypsin or other serine proteinases.) - A comparative study to probe the influence of biophysical effects on the BFs of such closely related residues (viz. CYS and SER), therefore, makes sense. While, SER is classified as a polar amino acid due to the presence of the hydroxyl group, CYS is considered polar mainly because of the perceived chemical analoguesness of its thiol group to the hydroxyl groups in the side-chains of other polar amino acids. But free CYS residues were reported (Nagano et al., 1999) to be present in hydrophobic regions of proteins and their hydrophobic nature can be observed to be almost equivalent to that of PHE in hydrophobic scale (Kyte and Doolittle, 1982).

- How does this piece of information reflect in the BF magnitude of CYS (as compared to SER)? Though with respect to structural consideration, the difference in mobility profile of CYS and SER should have been attributed only to the difference between local mobility of sulfur and oxygen atom, the effect of hydrophobic tendency of CYS (as compared to SER) can be observed in their comparative BF magnitudes. At the most global level, considering all the non-redundant proteins in 75 most populous folds, **Table-1** reports that SER BF is 25Å$^2$ as compared to CYS BF, 23Å$^2$.

Though small, this difference is significant because in **Table-1** the maximum-to-minimum variance of BF is merely ~9Å$^2$ (between GLU and VAL). One may reason that the lower BF of CYS is owing to its preference to be present in the (less mobile) hydrophobic regions of proteins; whereas SER, being a predominantly protein surface residue, is exposed to higher fluctuations which accounts for its higher BF (as compared to CYS). Such a hypothesis gathers consistent support from **Figure-1**, where it is found that SER BF is always higher than CYS BF irrespective of the structural class their host protein belongs to, viz. $\mathbf{SER_{BF}^{\alpha/\beta} - CYS_{BF}^{\alpha/\beta}} = 2.7$Å$^2$, $\mathbf{SER_{BF}^{\alpha+\beta} - CYS_{BF}^{\alpha+\beta}} = 1.1$Å$^2$, $\mathbf{SER_{BF}^{all-\alpha} - CYS_{BF}^{all-\alpha}} = 3.6$ Å$^2$, and $\mathbf{SER_{BF}^{all-\beta} - CYS_{BF}^{all-\beta}} = 0.2$Å$^2$. Since the sequence α/β→–α+β→–all-α→–all-β denotes the sequence of emergence of structural classes, one may also notice that protein evolution has treated SER consistently, whereby: $(\mathbf{SER_{BF}^{\alpha/\beta} - SER_{BF}^{\alpha+\beta}} (= 3.13$Å$^2)) \sim (\mathbf{SER_{BF}^{\alpha+\beta} - SER_{BF}^{all-\alpha}} (= 2.88$ Å$^2)) \sim (\mathbf{SER_{BF}^{all-\alpha} - SER_{BF}^{all-\beta}} (= 1.52$ Å$^2))$ - implying that while the trend to reduce SER mobility is kept steady, the extent of such mobility reduction itself is made to diminish uniformly during α/β→–α+β→–all-α→–all-β transitions. However, CYS BF evolution across structural classes failed to reveal any such uniform pattern, given by: $(\mathbf{CYS_{BF}^{\alpha/\beta} - CYS_{BF}^{\alpha+\beta}} (= 1.54$Å$^2)) \sim (\mathbf{CYS_{BF}^{\alpha+\beta} - CYS_{BF}^{all-\alpha}} (= 5.35$Å$^2)) \sim (\mathbf{CYS_{BF}^{all-\alpha} - CYS_{BF}^{all-\beta}} (= -1.87$Å$^2))$. While the reduction in residue mobility in α/β to α+β transition is recorded to be >2.2 Å$^2$ for most of the residues, extent of flexibility reduction for CYS during this transition is recorded to be the lowest, mere 1.54Å$^2$. On the other hand, while the reduction in residue mobility in α+β to all-α transition is recorded to be ~3.8Å$^2$ for most of the residues, extent of flexibility reduction for CYS during this transition is recorded to be huge 5.35Å$^2$, second only to that of LYS. While the reduction of residue-flexibility during all-α to all-β transition is found to be minimal or non-existent for most of the residues, CYS flexibility for this transition shows a decisively opposite trend whereby a significant increment of CYS BF could be registered during all-alpha to all-beta transition, which was found to be second only to, most interestingly, LYS. While standard deviation of SER and CYS (**Table-2**) were following the same trends during α/β→–α+β→–all-α transitions, increment in flexibility of CYS during all-α to all-β transition is reflected in its registering higher standard-deviation during it (as compared to that of SER). Furthermore, the breaking of monotonicity in CYS BF-profile caused it to register the poorest correlation coefficient (= -0.16) between evolutionary scale (Caetano-Anollés and Caetano-Anollés, 2003) and BF, as compared to that of SER (= -0.28). One may attempt to assign the cause behind observed absence of monotonicity in CYS BF profile and its poor correlation with evolutionary scale to the varied physico-chemical constraints that CYS was subjected to, which is presented in details in (Heitmann 1968; McConnell et al., 1993; Nagano et al., 1999; Rawlings et al., 2004). Effects of

CYS's having the lowest BF among all the polar residues on the biophysical properties of CYS can be estimated by the fact that even in the solvent-exposed highly flexible regions the loops are often stabilized by disulfide bonds, a process that demands maintenance of reduction of local mobility.

**3.2.2:** *The inverse problem of probing the finer details of biochemical properties of residues by analyzing their BFs.*

One can extend this perspective to view residue biophysical properties by studying their BF evolution patterns to other residues beyond CYS and SER. For example, side-chains of ILE, LEU and VAL are not reactive and thus are not involved in any covalent chemistry in enzyme active centers. Moreover they reside far from (mobile) solvent atoms. Not surprisingly therefore the mean and standard deviation of BFs of these residues (ILE:22.34(5.70), LEU:22.98(6.26), VAL:22.02(5.73)) contrast starkly with that of the reactive residues (GLU:31.25(7.29), ASP:28.93(6.81), ARG:28.20(7.22), LYS:28.84(8.02)) that commonly reside on the solvent-accessible surface of proteins. In fact upon clustering the residues with simple partition BF(mean)=25Å$^2$, we find two distinct classes with (ALA, VAL, LEU, ILE, PHE, TYR, TRP, CYS, MET, GLY) in one, registering mean BF<25Å$^2$ and (ARG, HIS, LYS, ASP, GLU, ASN, GLN, SER, PRO) in another, registering mean BF>25Å$^2$. Digging deeper, one finds that BF-based classification of the residues can be sensitive to distinguish between closely related residues too. For example, though VAL and THR are of roughly the same shape and volume and though it often proves to be difficult to distinguish VAL from THR (even in a high-resolution protein structure), merely because VAL differs from THR by replacement of the hydroxyl group with a methyl substituent, THR acquires slightly polar character; whereby, irrespective of structural classes THR registers higher BF than VAL (**THR$_{BF\text{-mean}}^{\alpha/\beta}$ − VAL$_{BF\text{-mean}}^{\alpha/\beta}$ = 2.20Å$^2$, THR$_{BF\text{-mean}}^{\alpha+\beta}$ − VAL$_{BF\text{-mean}}^{\alpha+\beta}$ = 2.72Å$^2$, THR$_{BF\text{-mean}}^{all\text{-}\alpha}$ − VAL$_{BF\text{-mean}}^{all\text{-}\alpha}$ = 1.87Å$^2$, THR$_{BF\text{-mean}}^{all\text{-}\beta}$ − VAL$_{BF\text{-mean}}^{all\text{-}\beta}$ = 1.59Å$^2$**). Attempting even more sensitive comparison, we consider the trends of BF difference between TYR and PHE across SCOP classes. PHE is nonpolar due to the hydrophobic nature of the benzyl side chain. TYR, although an aromatic amino acid like PHE, is derived from PHE by hydroxylation in the para position. Due to this very reason TYR becomes soluble which PHE is not. – Such tiny difference in biochemical properties between TYR and PHE accounts for minuscule difference in their BF profile, whereby not only the global BF of them register subtle but definite difference (**TYR$_{BF\text{-mean}}$ = 23.59Å$^2$, PHE$_{BF\text{-mean}}$ = 22.83Å$^2$**), but also the SCOP-class specific means (**TYR$_{BF\text{-mean}}^{\alpha/\beta}$ − PHE$_{BF\text{-mean}}^{\alpha/\beta}$ = 0.98Å$^2$, TYR$_{BF\text{-mean}}^{\alpha+\beta}$ − PHE$_{BF\text{-mean}}^{\alpha+\beta}$ = 0.49Å$^2$, TYR$_{BF\text{-mean}}^{all\text{-}\alpha}$ − PHE$_{BF\text{-mean}}^{all\text{-}\alpha}$ = 0.21Å$^2$, TYR$_{BF\text{-mean}}^{all\text{-}\beta}$ − PHE$_{BF\text{-mean}}^{all\text{-}\beta}$ = 1.10Å$^2$**). In a similar way one can attempt to compare B-factor profile of GLU and ASP.

With respect to ASP, GLU has one additional methylene group in its side chain. Due to the inductive effect of this additional methylene group the pKa of the GAMMA carboxyl group for GLU in a polypeptide measures about 4.3, higher than that of ASP (pKa of the β-carboxyl group of ASP in a polypeptide is about 4.0). Such subtle chemical differences account for subtle differences in GLU and ASP's reactivity profile also; whereby we observe that despite being subjected to similar evolutionary pressure (correlation coefficient between BF and evolutionary scale for ASP is -0.31, that for GLU is -0.32, details in **TABLE-3**) not only the global mean BF of GLU (31.25) is higher than that of ASP (28.93) but also, irrespective of structural classes GLU registered higher BF than ASP (**$GLU_{BF-mean}^{α/β} - ASP_{BF-mean}^{α/β}$ = 3.13Å$^2$, $GLU_{BF-mean}^{α+β} - ASP_{BF-mean}^{α+β}$ = 2.0Å$^2$, $GLU_{BF-mean}^{all-α} - ASP_{BF-mean}^{all-α}$ = 1.55Å$^2$, $GLU_{BF-mean}^{all-β} - ASP_{BF-mean}^{all-β}$ = 2.25Å$^2$**).

Hence one can generalize BF information to classify 20 residues with respect to their structural-class specific and evolution-specific mobility patterns. Such a perspective can be easily obtained by observing the general characteristics of **Figure-1**; whereby we find that:

1. for one group of residues (**class-I**, viz. VAL, THR, PHE, SER, PRO), the decrement of SCOP-class specific BF profile is observed to be strictly monotonous that is perfectly uniform along α/β→–α+β→–all-α→–all-β pathway;

2. for another group of residues (**class-II**, viz. ALA, MET, GLY, HIS, ARG, ASP), the decrement of SCOP-class specific BF profile is found to be strictly monotonous and uniform till α/β→–α+β→–all-α phase of structural class evolution; however, for all-α→–all-β transition the extent of decrement of BF slowed down;

3. the third group of residues (**class-III**, viz. ILE, LEU, TYR, ASN, GLU, GLN), similar trend of decrement of SCOP-class specific BF profile is strictly monotonous and uniform till α/β→–α+β→–all-α stage is observed; however, during all-α→–all-β transition the BF did not change at all;

4. finally, for the fourth group of residues (**class-IV**, viz. TRP, LYS, CYS), while adhering to the universal trend the decrement of SCOP-class specific BF profile shows strict monotonicity and uniformity till α/β→–α+β→–all-α stage, for all-α→–all-β transition this monotonicity is broken and during this transition mobility of these residues increased.

Such a classification of residues neatly explains the standard deviation(SD) patterns in their BFs. The SD of **class-I** residues reflected the monotonic and perfectly uniform decrement in their mean BF, whereby, as a result of the least variability trend of them the class-I residues registered the lowest SD (mean of **class-I** SDs = 6.22). The least tendency of variability from the trend of strict monotonicity coupled with uniform decrement in mean BF suggests clearly that BF decrement observed in VAL, THR, PHE, SER, PRO across all four major structural classes along the entire span of evolutionary journey has been most decisive. The fact that the trend of reduction of BF is not so decisive for **class-II** and **class-III** residues is reflected in their registering slightly higher SDs (mean of **class-II** SDs = 6.63, mean of **class-III** SDs = 6.49); implying that while the various biophysical constraints have dictated the reduction of their mobility, evolution hasn't been so decisive about reducing their flexibility.